# Fabrication and Characterization of Large Area Metallic Nano-Split-Ring Arrays by Nanoimprint Lithography

*Can Peng, Shufeng Bai and Stephen Y. Chou*

Nanostructure Laboratory, Princeton University, Princeton NJ 08544

AUTHOR EMAIL ADDRESS: canpeng@princeton.edu

ABSTRACT: This paper presents a novel method to parallel fabricate large area (wafer scale) metallic nano-split-ring arrays with nanoimprint lithography (NIL). To our knowledge it is the first method that can pattern large area and high dense metallic split-ring arrays with advantages of high throughput, low-cost and simplicity. This method makes metallic nano-split-ring arrays, which was somehow conceptual before, practically useful. The optical properties of the fabricated gold nano-split-ring arrays with different parameters were measured. They show very obvious magnetic response to the incident light (which shows 10dB extinction ration in transmission spectra). The structure fabricated by this method can generate magnetic response in optical range with relatively large feature size that relax the requirement of resolution on lithography.

I  INTRODUCTION

Metallic split ring arrays, which can generate magnetic response in optical frequencies, are key components for optical left-handed materials which can be used to build perfect lenses for nanoscale lithography [1, 2]. With the highly localized and enhanced optical field near this structure at plasmonic resonance wavelength they have many applications in bio and chemical sensors [3], and they work as optical antenna arrays which have many applications in nonlinear optics such as harmonic generation, mixing and signal processing [4,5]. In all the applications listed above the ability to fabricate large area of this structure with low-cost and high throughput is required to make the structure really practical useful.

Since split ring resonators were proposed as basic blocks for metamaterials with magnetic response, people have pushed hard to shrink the size of the split ring in order to bring the resonant frequency from microwave regime to optical regime [6, 7, 8]. A split ring resonator that has magnetic resonance in optical frequencies must have feature size in several-hundred-nanometer scale, in which range e-beam lithography is conventionally applied. The serial writing of the small structure with high density by e-beam is time-consuming and the writing area is limited to around 100 um$^2$. All these make large area (like wafer scale) fabrication almost impossible. In this work we proposed a novel parallel method to fabricate wafer scale nanoimprint molds of split ring arrays in nanometer size with low-cost and high throughput, which makes the metallic split-ring arrays practically useful in many areas.

The rest of this paper is organized as following: this novel fabrication method of split-ring-array molds is outlined and imprinted results of gold split-ring arrays are shown in section II, the discussion of the optical properties of these gold split-rings and the optical measurement results of these gold split-ring arrays with different parameters are presented in section III, section IV summarizes this paper.

## II   FABRICATION METHODS

The fabrication of split-ring molds is based on a fabrication method of square ring molds [6]. The whole process is depicted in Fig.1. This process starts from a 2D square pillar array mold made by double-imprint with a 1D grating mold, where the 1D grating mold was made by UV interference lithography [7].

First a square pillar array pattern is imprinted by using the square pillar mold on a Si wafer with a 100nm thermally oxidized layer on the top, then this pattern is transferred through the SiO2 layer to the Si wafer and a square SiO2 pillar array is obtained as shown in Fig.1 (a). The pitch of this array is around 1um and the size of each square pillar is around 400 by 400nm In order to make square ring mold from this pillar mold, a layer of SiNx with thickness of 120nm is conformally deposited on this SiO2 pillar array by liquid phase chemical vapor deposition LPCVD (Fig.1 (b)). After that the sample is anisotropic etched by reactive ion etching (RIE) to remove the SiNx layer on the top of square pillars (Fig.1 (c)) and let the SiO2 cores exposed, at the same time the SiNx layer on the Si substrate is also removed and SiNx ring walls around the SiO2 cores are left. Then the hydro fluoride (HF) solution is used to wash out the SiO2 cores in the SiNx rings and a square ring mold is left (Fig.1 (d)). By using UV nanoimprint a square ring array pattern is duplicated in the UV curable resist from the square ring mold we made and the pattern is transferred through the sublayer to the quartz wafer, then square trenches are formed in sublayer (Fig.1 (e)). In order to open gaps in the rings chromium (Cr) is shadow evaporated three time along the directions of the edges of the square rings (Fig.1 (f)) and the chromium patterns after each shadow evaporation are shown in the insert pictures, after lift-off Cr split-ring patterns are formed on the quartz wafer (Fig.1 (g)). After RIE etching a split-ring array mold is finally obtained (Fig.1 (h)). The size of the gaps in split-rings can be adjusted by tuning the angle of the shadow directions.

The SEM image of the 2D mother pillar array mold that was made by orthogonal double imprint from a 1D grating mold is shown in Fig.2 (a). And the SEM image of the square ring mold is shown in Fig.2 (b) and the SEM image of the Cr pattern after three times of shadow evaporations and lift-off is shown in Fig.2 (c). With the method described above very large size (we made mold on 4 inch wafer) mold can be fabricated. The SEM images at different positions over the 4 inch mold are shown in Fig.3 (a).

The uniformity of the split-ring mold over 4 inch wafer has been studied. The deviation of the ring size over the whole wafer is small (<3% in our case). This is due to the high uniformity of interference lithography during the grating mold fabrication (for 1 um grating over 4 inch wafer the deviation of pitch and duty cycle is around 0.2%). On the other hand the deviation of the split size is relatively larger (It is around 14% from the center to the rim of the 4 inch mold). There are several effects that affect the deviation of the split size, such as dispersion of the shadow angle, the uniformity of thickness of the resist and the wafer bending. The split size deviation is mainly caused by the dispersion of shadow angle (since the evaporated Cr atom beam is not really collimated) during the shadow evaporation of Cr, as shown in Fig.3 (b). From the geometry of evaporator setup it can be seen that small shadow angle, $\theta_0$ (the shadow angle at center defined in Fig.3 (b)), will cause large deviation of split size, thus large shadow angle is preferred. The largest deviation of split size over the whole wafer can be expressed as $1/{k \sin \theta_0}$, where k is the ration of the distance between Cr source to wafer to the wafer radius. In order to keep certain mean split size the thickness of the resist (depth of the trenches) should be reduced under large shadow angle. The uniformities of two molds with almost the same mean split sizes are compared. During fabrication process of these two molds different shadow angles ($\theta_0=45°$ and $\theta_0=30°$ respectively) are used to open the splits. It can be seen that the uniformity of the mold that was fabricated with a larger shadow angle (45°) is better compared with the other as shown in Fig.3 (c). So we can see that the large shadow angle in the third Cr evaporation process is preferred to achieve high uniformity of split size over the wafer. In order to keep a certain mean split size the thickness of resist should small under large shadow angle.

Due to the nanometer resolution of nanoimprint lithography, the method described above puts little limitation on the feature size. The only limitation comes from the 1D grating mother mold. Since with UV interference lithography 1D grating mold with pitch around 150nm can be made, our method can be used to fabricate sub 100nm split-ring arrays with extremely high density.

Another thing needed to be point out is that mold for split rings with three splits can be fabricated by changing shadow evaporation process. If two times (the first time and the third time in the shadow evaporation process discussed above) instead of three times of shadow evaporation are applied, three splits can be opened in the evaporated Cr patterns.

III  OPTICAL PROPERTIES

Three possible plasmonic modes that can exist in the gold split-ring are shown in Fig.4. Two of them (Fig.4 a, b) are defined as electrical modes since they can only generate electrical response to the incident light. In those cases the collective oscillations of electrons in the two pieces of the gold split-ring are in phase respect to each other, and thus no current loop, which is responsible to the magnetic response to the incident light, exist in those modes.

The magnetic response of the metallic split-rings to normally incident light can be generated when the plasmonic resonances of the electrons in the two pieces of the gold split-ring are in reversed phase (Fig.4 c), which is due to the gaps opened in the rings. With the gaps a split ring can be viewed as a miniature LC circuit resonator. The gaps can be viewed as capacitors and the rings as inductor. So the magnetic resonance frequency can be expressed as resonance frequency of a LC resonator, $\omega = \sqrt{2/LC}$. The numerator of 2 in the square root is due to that there are two gaps in the gold split ring. This shows another merit of this structure --- blue shift of the magnetic resonance frequency by $\sqrt{2}$. That means resonance frequency in optical range can be obtained with gold split-rings with relatively larger size compared with single-gap split-rings. Compared with the structure in Ref. 8, our structure has doubled feature size but almost the same magnetic resonance frequency as shown in following.

From the symmetry of the structure fabricated by the method described in previous section, we can see that the magnetic response should be polarization-dependent. The square split-ring has two gaps in x-direction as shown in Fig.3, thus it is symmetric about x-axis. Meanwhile the plasmonic mode in this structure which can introduce magnetic response by generating loop current has anti-symmetry about x-axis. Then the symmetric metallic structure can not mediate coupling between the anti-symmetric plasmonic mode and the normally incident x-direction polarized light which is symmetric about x-axis. Because the split-ring and the magnetic plasmonic mode both are asymmetric about y-axis, the coupling between the normally incident y-direction polarized light and the magnetic plasmonic mode in the structure is possible.

The coupling between the incident light and the plasmonic modes in the split-rings can be simulated by using discrete dipole approximation (DDA) [11,12]. By this method the extinction cross-section spectrum of the split-ring can be calculated. When the metallic structure is resonating under the exciting of the incident light at certain frequency, the extinction cross-section of the structure at that frequency is enhanced tremendously. The calculated cross-section spectra of a single gold split-ring are shown in Fig.4. In Fig.4 (d) the extinction cross-section spectra of a split ring (the outer length of each side is 700nm and the width of the ring is 160nm and the thickness of the ring is 50nm) under the two orthogonal polarizations of the normally incident light, as a reference the extinction cross-section spectra of gold ring without split also shown there. Due to the symmetry rules in coupling, when the polarization is along the symmetric axis, only an electric mode can be excited in the structures as the spectrum shows, and two resonance peaks can be seen when the polarization is rotated by 90° to along the asymmetric axis of the structure. The extinction cross-section spectra of the split-rings with different split sizes under polarization along asymmetric are shown in Fig.4 (e). The resonance wavelength of the magnetic mode shifts to blue side as the size of the splits increases.

The transmission spectra of the fabricated gold split-rings arrays with different gap sizes were measured by using Fourier transform infrared spectrometer (Nicolet 730 FT-IR Spectrometer), as shown in Fig.5. The fabricated gold split-rings are in square shape. The length of each side is around 790nm and the frame width of these split square rings is around 160nm. Split-rings with three different sizes of gaps were fabricated: 0nm (no gap), 44nm, 144nm and 180nm. All samples use

Pyrex glass as substrates. Gold thickness is around 50nm and a thin layer of Ti (3nm) was deposited as adhesion layer between substrates and the gold layer. In Fig.5 (a) the transmission spectra of the gold ring array without gap under two orthogonal polarizations are presented as references. As expected there is only one dip around 2.62μm in the observed range, which is corresponding to an electric resonance mode. And the optical response of the whole rings is polarization-independent. The transmission spectra of gold split-rings with gap size of 44nm under two orthogonal polarizations are shown in Fig.5 (b). The characteristics of these spectra are well consistent with theoretical prediction. When the polarization direction is along the symmetric axis of this structure, there is only one resonance observed around 1.72μm in the wavelength range measured, which is corresponding to an electric mode. As discussed above under this polarization no magnetic mode can be excited due to the symmetry of the system. Then we turned the polarization by 90 degree along the direction of the axis about which the structure is asymmetric and under this configuration obvious extra resonances (extinction ration is around 10dB) can be observed around 3.55μm and 1.55μm in the transmission spectra, which are corresponding to magnetic modes. The resonance around 1.55μm, which can not be predicted by response of single split-ring, is caused by the coupling between plamonic modes of adjacent split-ring in the array. In order to further prove that those extra resonances are magnetic ones we measured the transmission spectra under this configuration (incident light polarized along asymmetric axis) of gold split-ring arrays with different gap sizes, as shown in Fig.5 (c). With increasing of gap size the resonance wavelengths of the extra modes have blue-shift. Since the large gap size means large capacitance in the equivalent LC circuit, increasing of the gap size leads to short wavelength of the LC resonance the equivalent of the magnetic plasmonic resonance in the split-ring. This well explains the blue-shift of the resonance wavelength of those extra modes. It confirms that the extra resonance dips in transmission spectra are corresponding to magnetic modes. The plasmonic mode that has resonance wavelength around 2.15μm independent of the gap sizes is obviously electric mode. This wavelength is determined by the size (790nm) of gold pieces along the asymmetric axis. Fig.5 (d) shows the transmission spectra of the samples with different gap sizes when the incident light is polarized along the symmetric axis of the structure. For this polarization there are only electric modes. The resonance wavelengths are determined by sizes of gold pieces along the direction of the polarization. Thus large gaps lead to small sizes along the polarization direction and then lead to short resonance wavelength as we can observe in Fig.5 (d).

We also investigated the effective magnetic permeability of the split ring array when the magnetic field is perpendicular to the plane defined by the split rings as shown in Fig.6(a). The effective magnetic permeability can be expressed as [13]

$$\mu_{eff} = 1 - \frac{F\omega^2}{\omega^2 - \frac{1}{CL} + i\frac{Z(\omega)\omega}{L}}$$

where F is the filling factor of the array, C and L are geometrical capacitance and inductance of the split ring. Z is the impedance of the material of which the split ring is made. The inductance of the split ring can be calculated by the method given in [14]. The calculated spectrum of effective magnetic permeability is shown in Fig.6(b). It can be seen that around the resonance peak the effective permeability can be dramatically deviated from unity. Especially within a wavelength range that is smaller than the resonance wavelength, the effective permeability can be negative.

IV CONCLUSIONS

In conclusion we first proposed a novel fabrication method that is suitable for patterning metallic split-ring arrays with high density over wafer scale area. This fabrication method is purely a parallel process, and thank to the advantages of nanoimprint lithography it can be applied to pattern hundreds nanometer scale split-ring array with high throughput and low cost. All these properties of this method make metallic split-ring arrays practically useful in many areas, such as superlenses, optical magnetic metamaterials, bio/chemical sensors, nonlinear optics and etc.

With this method we fabricated gold nano split-ring arrays with different gap sizes. The physics of the plasmonic resonating modes in these structures was discussed. The optical response of these structures was measured. Obvious magnetic response of the gold split-rings was observed, and the behavior of the magnetic modes was well consistent with theoretical prediction.

ACKNOWLEDGMENT

Authors would like to thank the supporting from

FIGURE CAPTIONS

Fig.1. The fabrication process of large area nano split-ring molds. (a) fabricate a 2D square SiO2 pillar array on Si substrate by nanoimprint, the large area mother 2D pillar mold was made by double nanoimprint from a 1D grating mold made by interference lithography. (b) conformally deposit a layer of SiNx of 120nm by using LPCVD. (c) etch the SiNx layer on the top of pillars and the residual layer of SiNx on the substrate by RIE, and SiNx square rings around SiO2 cores are left. (d) etch the SiO2 cores with HF solution, and SiNx rings are left. (e) use the SiNx ring array as mold to imprint square trenches in imprint resist, and transfer patterns to substrate. (f) three times of shadow evaporations of Cr are done from three different angles, and the insert shows the metallic patterns on the substrate after each of the shadow evaporations. (g) do lift-off and a metallic split-ring array pattern is left on the substrate. (h) etch down the pattern into substrate and remove the mask by CR-7 and split-ring mold is made.

Fig.2 (a) the 2D pillar array mother mold made by orthogonal double imprint from a 1D grating mold that is fabricated by interference lithography. (b) the SiNx ring array mold made by the process described in Fig.1. (c) a ring trench array in resist imprinted by the SiNx ring array mold. (d) the Cr pattern left on the substrate after three times of shadow evaporations and lift-off.

Fig.3 (a) SEM images of the split ring mold at different positions over the 4 inch wafer. The deviation of the split sizes at the bottom and top of the wafer from that at the center are relatively large. That is caused by different shadow angles as shown in the setup of evaporator in part (b). (b) the setup scheme of the evaporator we used. Since the distance between the Cr source and the wafer is limited (around 0.5 m), the shadow angles are different at different places on the wafer. As shown the shadow angle at the center of the wafer is $\theta_0$, and those are $\theta_1$ and $\theta_2$ at the bottom and the top of the wafer respectively. (c) the uniformities of molds from bottoms to tops, which are fabricated under different shadow angles during the Cr evaporations. It can be seen that the uniformity of the mold made under $45^o$ shadow angle is better than that of the mode made under $30^o$ shadow angle.

Fig.4. (a) fabricated gold split ring arrays with different split sizes. (b), (c) and (d) show possible plasmonic modes in a gold split ring. (b) an electric mode in which case collective electron oscillations in the two pieces of the gold split-ring are in-phase along the symmetric axis of the structure. (c) another electric mode in which case collective electron oscillations in the two pieces of the gold split-ring are in-phase along the asymmetric axis of the structure. (d) the magnetic mode in which case collective electron oscillations in the two pieces of the gold split-ring are in opposite directions and generate a loop current that can introduce magnetic response to incident light. And the gold split-ring under this magnetic mold can be simplified as a LC resonator. (e) the extinction cross-section spectra simulated by DDA for gold split ring under normally incident light of two orthogonal polarizations are shown, and as a reference that of a gold ring without split is also shown. (f) the extinction cross-section spectra of gold split-rings with different split sizes under normally incident light polarized along the asymmetric axis are shown and the blue shift of one extinction peak with the increase of the split size indicate that this peak is corresponding to the magnetic mode.

Fig.5. (a) measured transmission spectra of a gold ring array without split under normally incident light of two orthogonal polarizations are shown as references. (b) measured transmission spectra of a gold split-ring array with split size of 44nm under normally incident light of two orthogonal polarizations are shown. (c) measured transmission spectra of gold split-ring arrays with different split sizes of 44nm, 144nm and 180nm under normally incident light polarized along the asymmetric axis of these structures. (d) measured transmission spectra of gold split-ring arrays with different split sizes under normally incident light polarized along the asymmetric axis of these structures.

Fig.6 Calculated effective permeability of split ring arrays. (a) the configuration of incidence of light. (b) the effective permeability spectrum. The parameters are chosen as a=790nm, w=160nm, g=144nm and the thickness of the whole structure is 50nm.

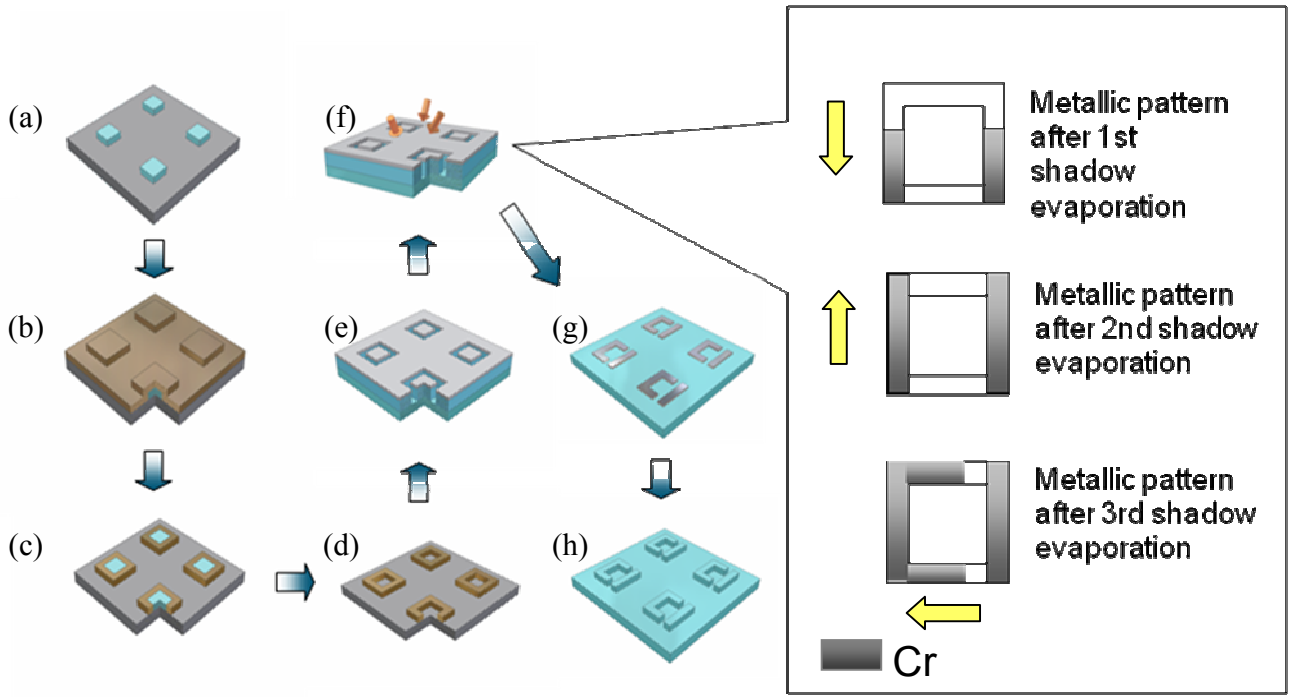

Fig.1

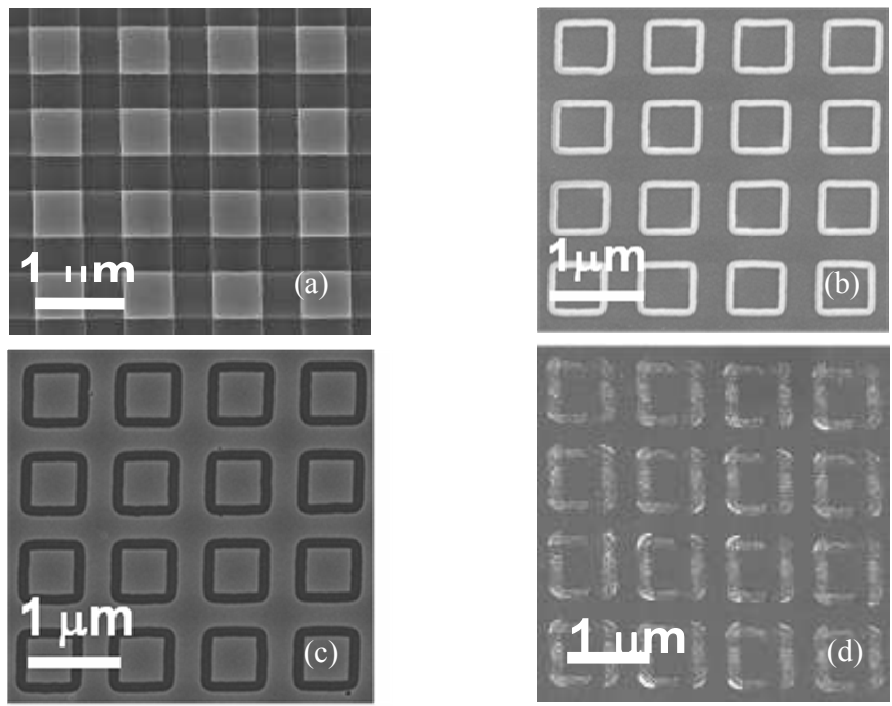

Fig.2

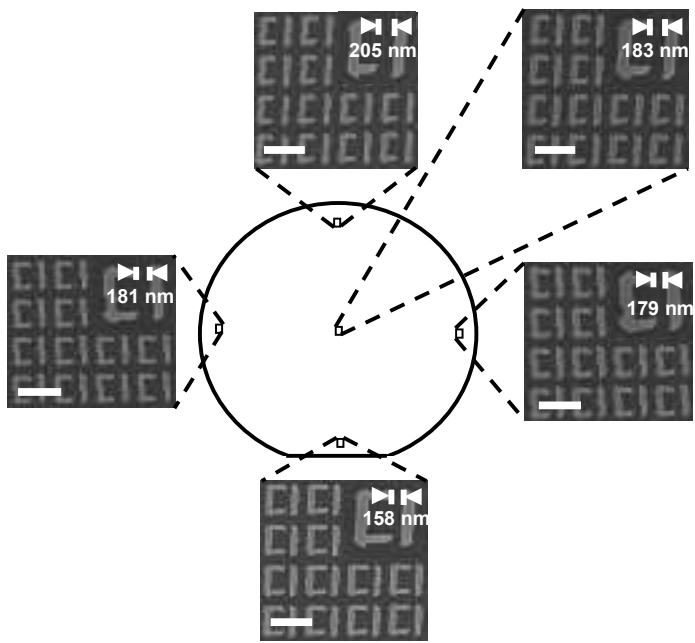

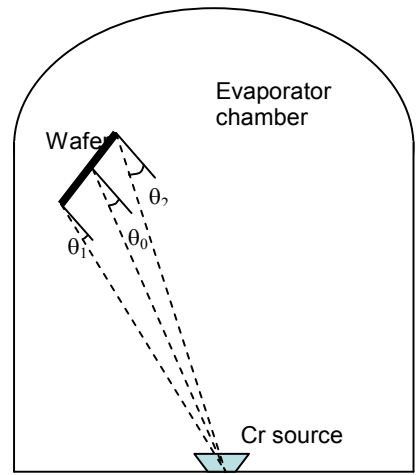

(a)

(b)

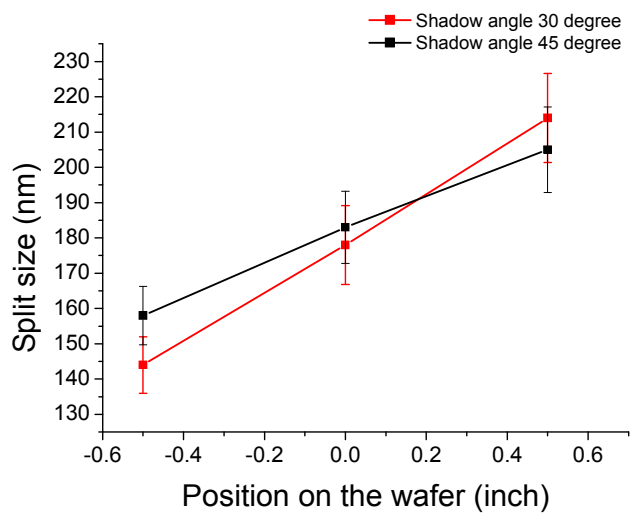

(c)

Fig.3

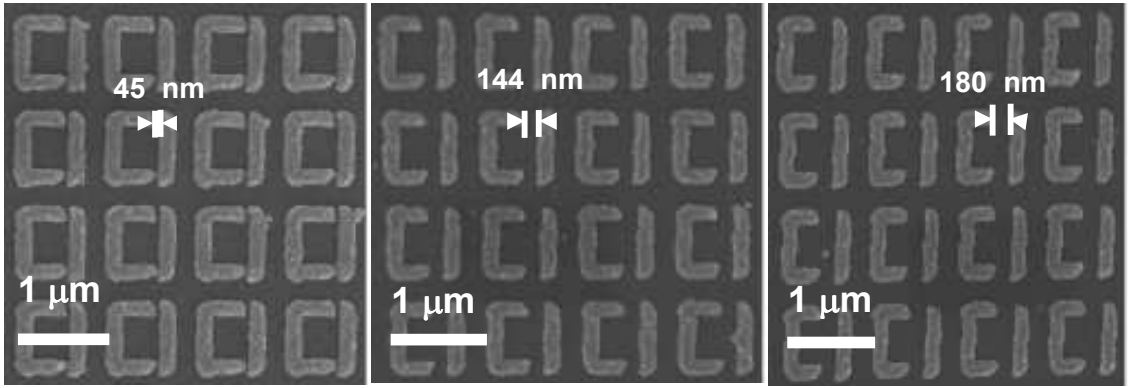

(a)

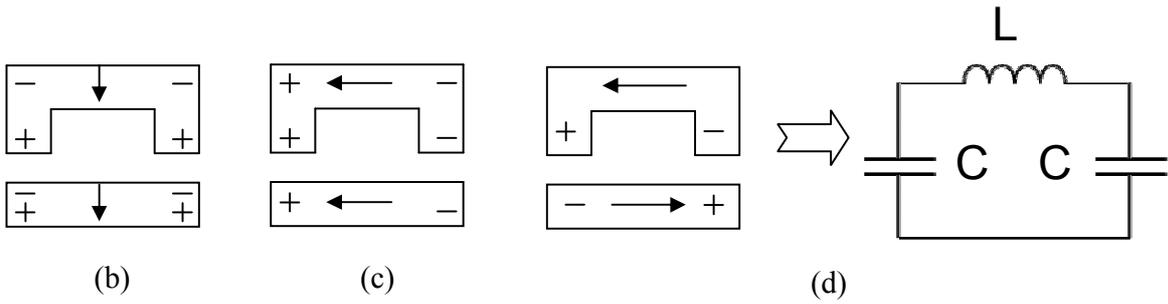

(b)　　　　　(c)　　　　　　　(d)

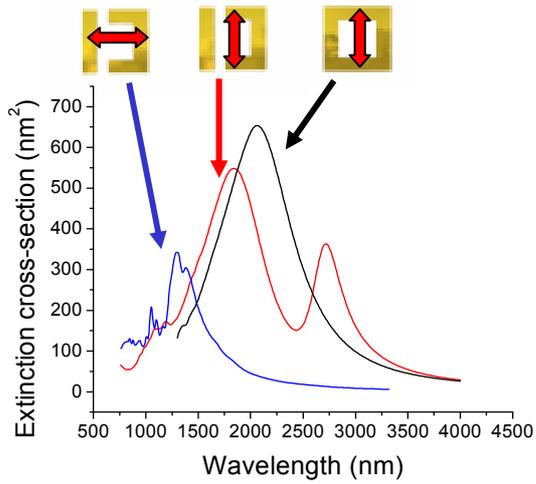

(e)

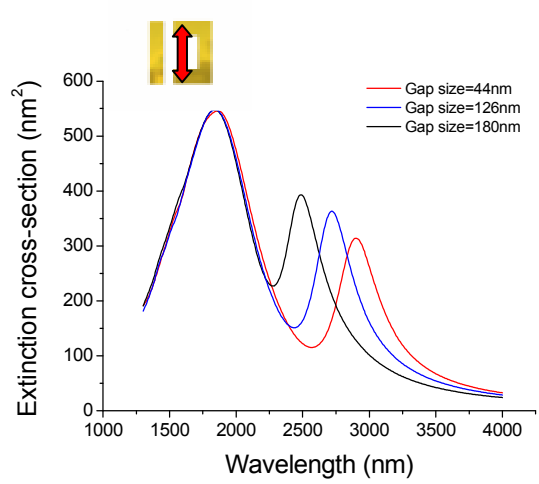

(f)

Fig. 4

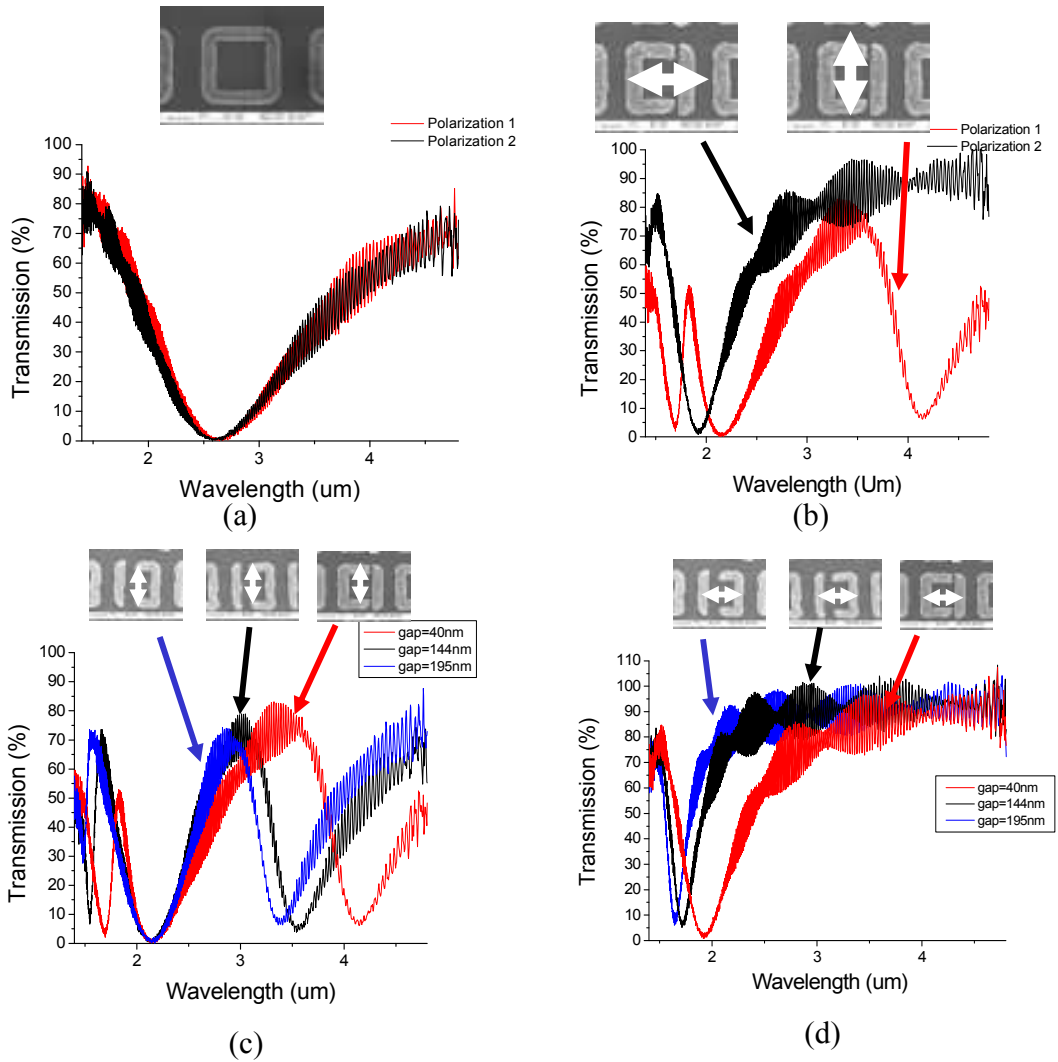

Fig.5

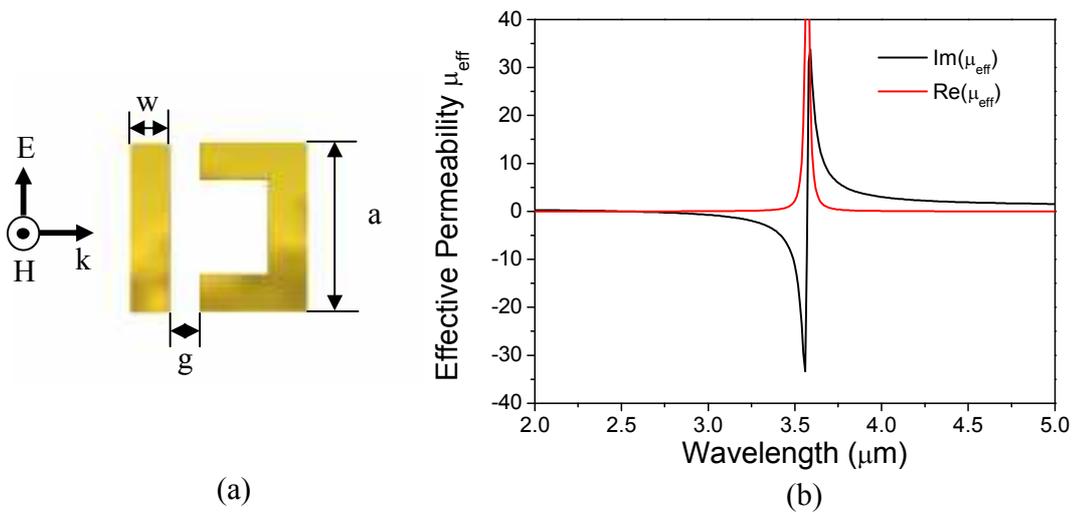

Fig.6